\RequirePackage[l2tabu, orthodox]{nag}
\RequirePackage{snapshot}

\documentclass[10pt,onecolumn]{article}

\makeatletter
\if@twocolumn
  \usepackage[dvips,letterpaper,top=0.5in, bottom=0.5in, left=0.75in, right=0.5in,includefoot,heightrounded]{geometry}
\else
  \usepackage[dvips,letterpaper,margin=1in,includefoot,heightrounded]{geometry}
\fi

\usepackage{srcltx}

\usepackage{amsmath}
\usepackage{amssymb,amsfonts}

\usepackage{abstract}

\usepackage[usenames,dvipsnames]{color}
\usepackage[usenames,dvipsnames]{xcolor}
\usepackage{subfigure}

\usepackage{booktabs}

\usepackage{setspace}
\usepackage{flushend}
\usepackage{multicol}

\usepackage{cite}
\usepackage{url}
\usepackage[normalem]{ulem}

\usepackage{enumerate}

\usepackage{hyperref}

\usepackage[sc,tiny,center]{titlesec}
       \usepackage{mathptmx}
\usepackage{graphicx}



\title{%
Multilayer Hadamard Decomposition of the Discrete Hartley Transform
}

\author{%
H. M. de Oliveira%
\thanks{%
H. M. de Oliveira
was with the
Grupo de Pesquisa em Comunica\c{c}\~oes (CODEC),
Departamento de Eletr\^onica e Sistemas,
Universidade Federal de Pernambuco.
Currently he is with the
Signal Processing Group,
Departamento de Estat\'{\i}stica,
Universidade Federal de Pernambuco.
Email:~\url{hmo@ufpe.br}
}
\and
R.~J.~Cintra%
\thanks{%
R.~J.~Cintra
was
with the Graduate Program in Electrical Engineering,
Universidade Federal de Pernambuco.
Currently he is
with
the Signal Processing Group,
Departamento de Estat\'{\i}stica,
Universidade Federal de Pernambuco.
E-mail:~\url{rjdsc@de.ufpe.br}
}
\and
R. M. Campello de Souza%
\thanks{%
R. M. Campello de Souza
is with the
Departamento de Eletr\^onica e Sistemas,
Universidade Federal de Pernambuco.
Email:~\url{ricardo@ufpe.br}
}
}

\date{}

\newcommand{\myabstract}{%
Discrete transforms such as the discrete Fourier transform (DFT) or the discrete Hartley transform (DHT) furnish an indispensable tool in signal processing.
The successful application of transform techniques relies on the existence of the so-called fast transforms.
In this paper some fast algorithms are derived which meet the lower bound on the multiplicative complexity of
the DFT/DHT.
The approach is based on a decomposition of the DHT into layers of Walsh-Hadamard transforms.
In particular,
fast algorithms
for short block lengths such as $N \in \{4, 8, 12, 24\}$ are presented.
}

\newcommand{\mykeywords}{%
Hadamard transform, discrete Hartley transform, fast algorithms
}

\begin{document}

\makeatletter
\if@twocolumn

\twocolumn[%
  \maketitle
  \begin{onecolabstract}
    \myabstract
  \end{onecolabstract}
  \begin{center}
    \small
    \textbf{Keywords}
    \\\medskip
    \mykeywords
  \end{center}
  \bigskip
]
\saythanks

\else

  \maketitle
  \begin{abstract}
    \myabstract
  \end{abstract}
  \begin{center}
    \small
    \textbf{Keywords}
    \\\medskip
    \mykeywords
  \end{center}
  \bigskip
  \onehalfspacing
\fi

\section{Introduction}

Discrete transforms defined over finite or infinite fields have been playing a relevant role in 
numerical analysis. 
A striking example is the discrete Fourier transform (DFT), which has found applications in several areas, especially in signal processing. 
Another relevant example concerns the discrete Hartley transform (DHT)~\cite{ref-1}, the discrete version of the integral transform introduced by Hartley in~\cite{ref-2}.
Besides its numerical side appropriateness, the DHT has proven over the years to be a powerful tool~\cite{ref-3,ref-4,ref-5}.
A decisive factor for applications of the DFT has been the existence of the so-called fast transforms (FT) for computing it~\cite{ref-6}.
Fast Hartley transforms also exist and are deeply connected to the DHT applications~\cite{ref-7,ref-8}.
Recent promising applications of discrete transforms concern the use of finite field Hartley transforms~\cite{ref-9} to design digital multiplex systems, efficient multiple access systems~\cite{ref-10} and multilevel spread spectrum sequences~\cite{ref-11}.

Fast algorithms that present low multiplicative complexity
are of relevant interest to community.
Very efficient algorithms such as the prime factor algorithm (PFA) or Winograd Fourier transform algorithm (WFTA) have also been used~\cite{ref-13,ref-14}.
Another particular class of algorithms that aims at low multiplicative complexity is
the arithmetic Fourier transforms (AFT)~\cite{ref-12}.
The minimal multiplicative complexity, $\mu$, of the one-dimensional DFT for all possible sequence lengths, $N$, can be computed by converting the DFT into a set of multi-dimensional cyclic convolutions. A lower bound on the multiplicative complexity of the DFT is given in~\cite[Theorem 5.4, p.~98]{ref-15}. For some short blocklengths, the values of $\mu(N)$ are given in Table~\ref{table-1} (some local minima of $\mu$).
The discrete Hartley transform of a signal $v_i$,
$i=0,1,2,\ldots,N-1$
is defined by:
\begin{align*}
V_k
=
\sum_{i=0}^{N-1}
v_i
\cdot
\operatorname{cas}
\left(
\frac{2\pi ik}{N}
\right)
,
\quad
k=0,1,\ldots,N-1
.
\end{align*}
where
$\operatorname{cas}(x)=\cos(x)+\sin(x)$
is the ``cosine and sine'' Hartley symmetric kernel.

\begin{table}
\centering
\caption{Minimal multiplicative complexity for computing the DFT of length $N$}
\label{table-1}
\begin{tabular}{cc}
\toprule
$N$ & $\mu(N)$ \\
\midrule
4 & 0 \\
8 & 2 \\
12 & 4 \\
24 & 12\\
\bottomrule
\end{tabular}
\end{table}

In this paper, 
we aim at the introduction of fast algorithms 
that meet the minimal multiplicative complexity.
There is a simple relationship between the DHT and the DFT spectra
of a given real discrete signal 
$\mathbf{v} = \begin{bmatrix}v_0&v_1&\cdots&v_{N-1}\end{bmatrix}^\top$,
$i=0,1,\ldots,N-1$.
Let 
$\begin{bmatrix}U_0 & U_1 & \cdots & U_{N-1}\end{bmatrix}^\top$
and
$\begin{bmatrix}V_0 & V_1 & \cdots & V_{N-1}\end{bmatrix}^\top$
be
the
DFT and DHT spectra
of~$\mathbf{v}$,
respectively.
Then,
we have that
\begin{align*}
V_k
&
=
\Re\{U_k\}
-
\Im\{U_k\}
,
\\
U_k
&
=
\frac{V_k + V_{N-k} - j \cdot (V_k - V_{N-k})}
{2}
.
\end{align*}
Therefore, an FFT algorithm for the DHT is also an FFT for the DFT and vice-versa~\cite[Corollary~6.9]{ref-15}.
Besides being a real transform, the DHT is also an involution, i.e., the kernel of the inverse transform is exactly the same as the one of the direct transform (self-inverse transform).
We exploit the DHT symmetry to derive 
fast algorithms that attain the theoretical minimal number of real floating-point multiplications.
The idea behind our approach is to carry out the DHT decomposition based on classical transforms by Hadamard~\cite{ref-16}.

In this work,
we adopt the following notation.
The input signal is denoted as~$\mathbf{v}$.
The DHT spectrum of~$\mathbf{v}$ 
is~$\mathbf{V} = \begin{bmatrix}V_0 & V_1 & \cdots & V_{N-1}\end{bmatrix}^\top$.
The DHT matrix of size~$N$ is referred to as $\mathbf{H}_N$
whose $(i,k)$th entry
is given by
$h_{i,k} = 
\operatorname{cas} \left( \frac{2\pi (i-1)\cdot(k-1)}{N} \right)$,
$i,k=1,2,\ldots,N$

\section{Computing the 4-point DHT}

For $N=4$,
we have the 
matrix formulation 
$\mathbf{V}=\mathbf{T}_4 \cdot \mathbf{v}$,
which is given by
\begin{equation*}
\begin{split}
\includegraphics[width=4cm]{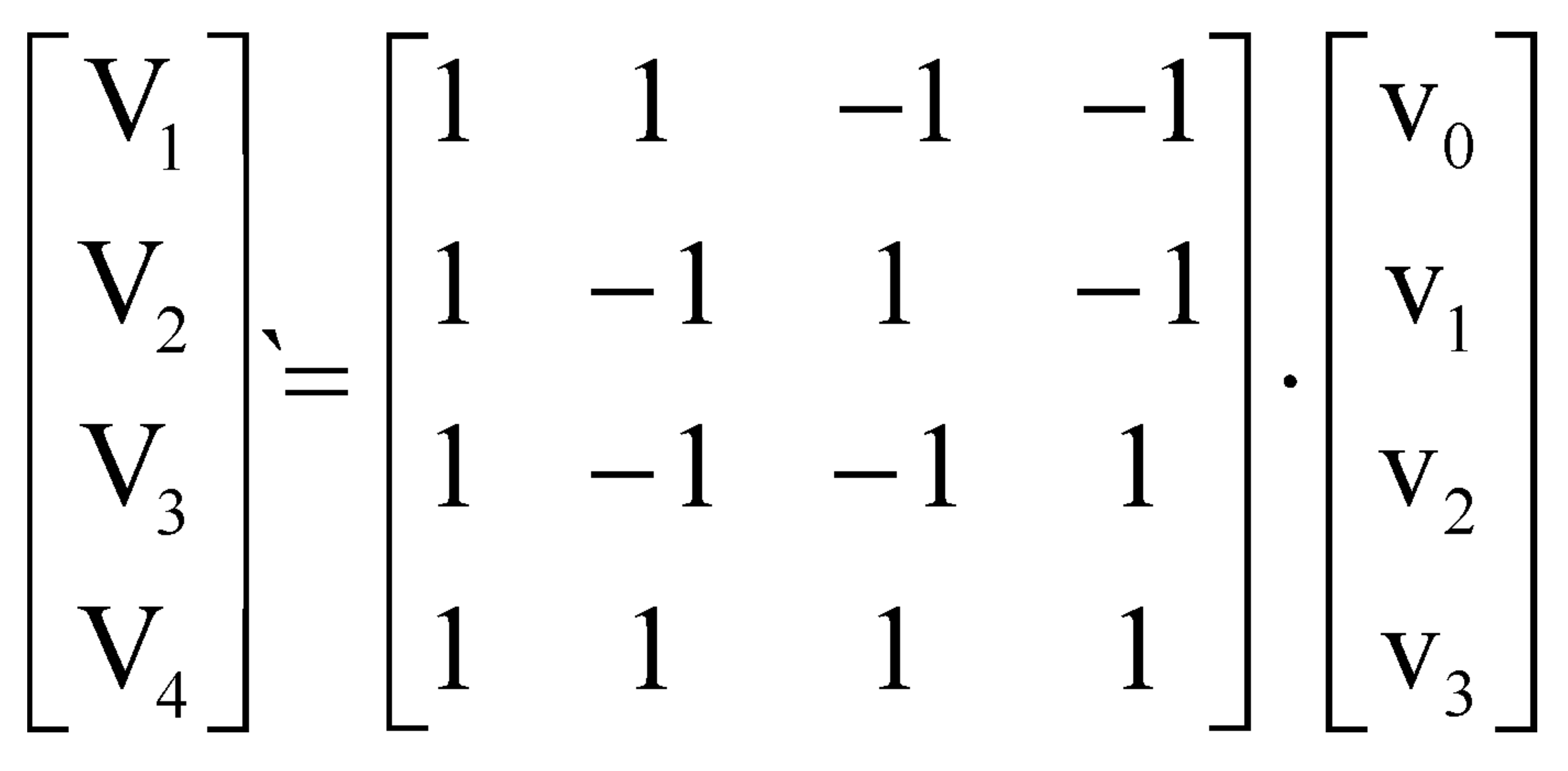}
\end{split}
.
\end{equation*}

\begin{figure}
\centering
\subfigure[]{\includegraphics[]{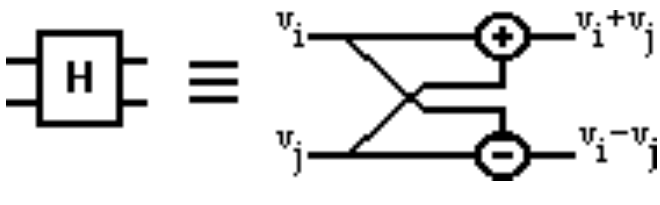}}
\qquad
\subfigure[]{\includegraphics[width=5cm]{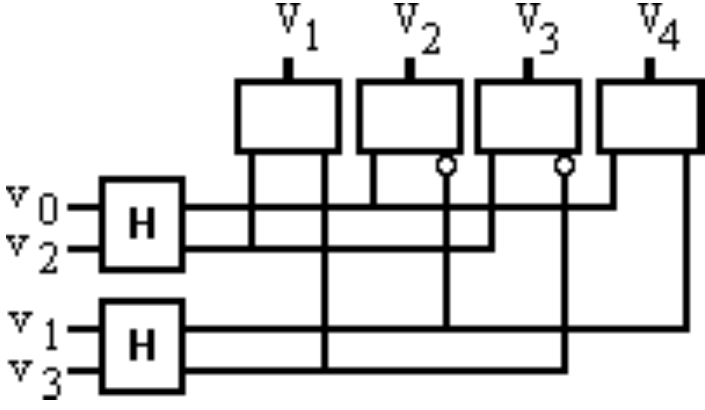}}

\caption{%
(a)~Diagram for the 2-point Walsh-Hadamard transform
and
(b)~Diagram for the 4-point
DHT based on Walsh-Hadamard transform. 
Small circles at the summation boxes indicate the subtraction operation
(invert the sign of the input) and
the ``H'' blocks denote the Hadamard transform.
}
\label{figure-1}
\end{figure}

It is therefore equivalent to the 4-point Walsh-Hadamard transform.
Thus it has null multiplicative complexity.
Figure~\ref{figure-1} shows the signal flow diagram
of the 4-point DHT in terms of 2-point Walsh-Hadamard transforms.
The complexity for the 4-DHT is given by 8~additions and zero multiplications.

\section{Computing the 8-point DHT}

Let $S_i(0)=v_i$, $i=0,1,\ldots,7$ (input data).
The 0-order ``pre-additions'' are, respectively, 
$\{S_0(0)=v_0, S_1(0)=v_1, S_2(0)=v_2, S_3(0)=v_3, S_4(0)=v_4, S_5(0)=v_5, S_6(0)=v_6, S_7(0)=v_7 \}$.
Thus,
8-point DHT matrix can be written as:
\begin{equation*}
\begin{split}
\includegraphics[width=8cm]{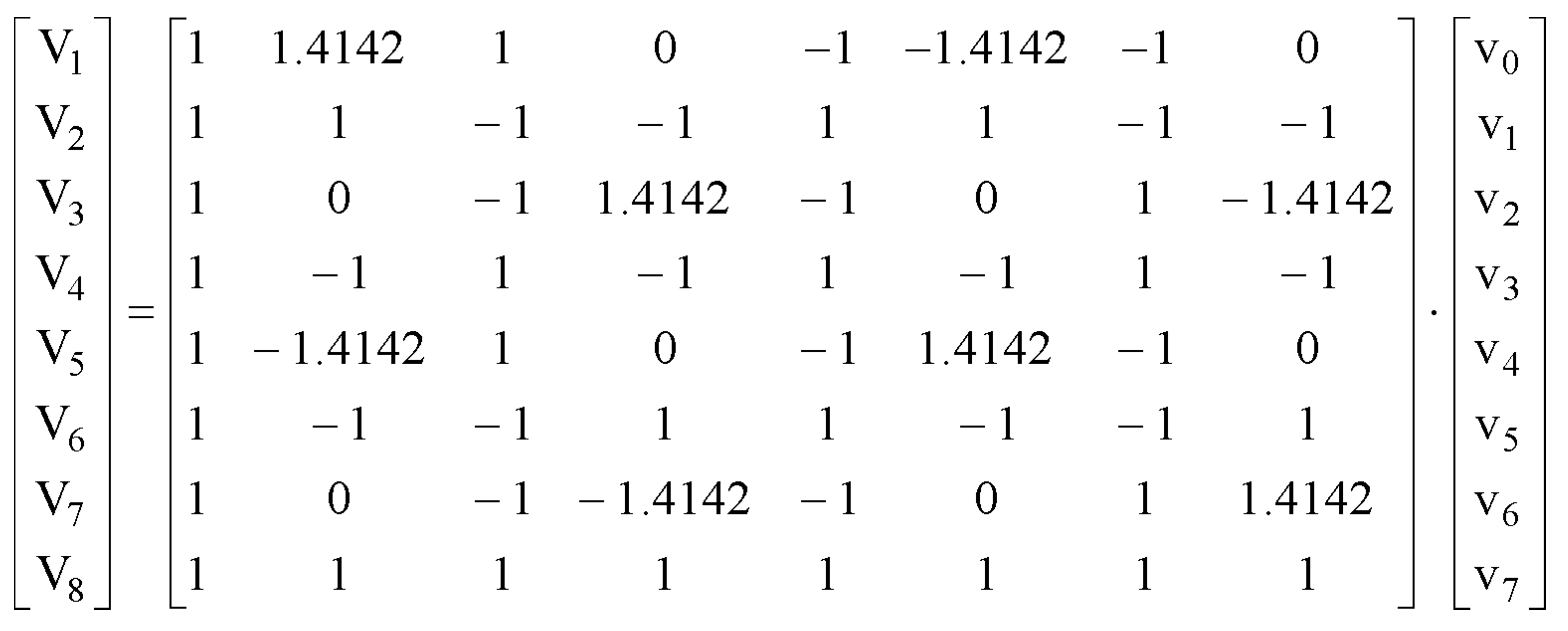}
\end{split}
.
\end{equation*}

We remark that 
\begin{align*}
\operatorname{cas}\left( \frac{2\pi k (i+N/2)}{N} \right)
=
\operatorname{cas}\left( \frac{2\pi k i}{N} + \pi k \right)
=
(-1)^k
\cdot
\operatorname{cas}\left( \frac{2\pi k i}{N}\right)
,
\end{align*}
which follows from the addition of arcs formula: 
$\operatorname{cas}(\alpha-\beta)=\cos(\beta) \cdot \operatorname{cas}(\alpha)- \sin(\beta) \cdot \operatorname{cas}'(\alpha)$, where 
$\operatorname{cas}'(\cdot)$
is the complementary cas function
$\operatorname{cas}'(\alpha)=\cos(\alpha)-\sin(\alpha)$ [3]. 
We notice that
the absolute value of the elements of the 2nd column
are identical to the corresponding elements at the 6th column; the same for the 3th column and 7th column.
We can thus consider new variables
$(v_1+v_5)$ and $(v_1-v_5)$ instead of $v_1$ and $v_5$; 
$(v_2+v_6)$ and $(v_2-v_6)$ instead of $v_2$ and $v_6$, and so on.
Thus, 
we obtain:
\begin{align*}
S_0(1)&=(v_0+v_4), \qquad S_1(1)=(v_0-v_4),
\\
S_2(1)&=(v_2+v_6), \qquad S_3(1)=(v_2-v_6),
\\
S_4(1)&=(v_1+v_5), \qquad S_5(1)=(v_1-v_5),
\\
S_6(1)&=(v_3+v_7), \qquad S_7(1)=(v_3-v_7)
.
\end{align*}
We refer to the above set of equations as 
the 1st-order pre-additions.
The first-order pre-additions effects several null elements in 
the implied new transform matrix. 
Although such an implementation requires only two multiplications, 
we may go further and combine other columns,
resulting in a alternative
2nd-order pre-additions
as follows:
\begin{align*}
S_0(2)&=(v_0+v_4), \qquad S_1(2)=(v_0-v_4), \\
S_2(2)&=(v_2+v_6), \qquad S_3(2)= (v_2-v_6), \\
S_4(2)&=(v_1+v_5)+(v_3+v_7), \qquad S_5(2)=(v_1+v_5)-(v_3+v_7), \\
S_6(2)&=(v_1-v_5)+(v_3-v_7), \qquad S_7(2)=(v_1-v_5)-(v_3-v_7)
.
\end{align*}

Thus,
we have:
\begin{equation*}
\begin{split}
\includegraphics[width=8cm]{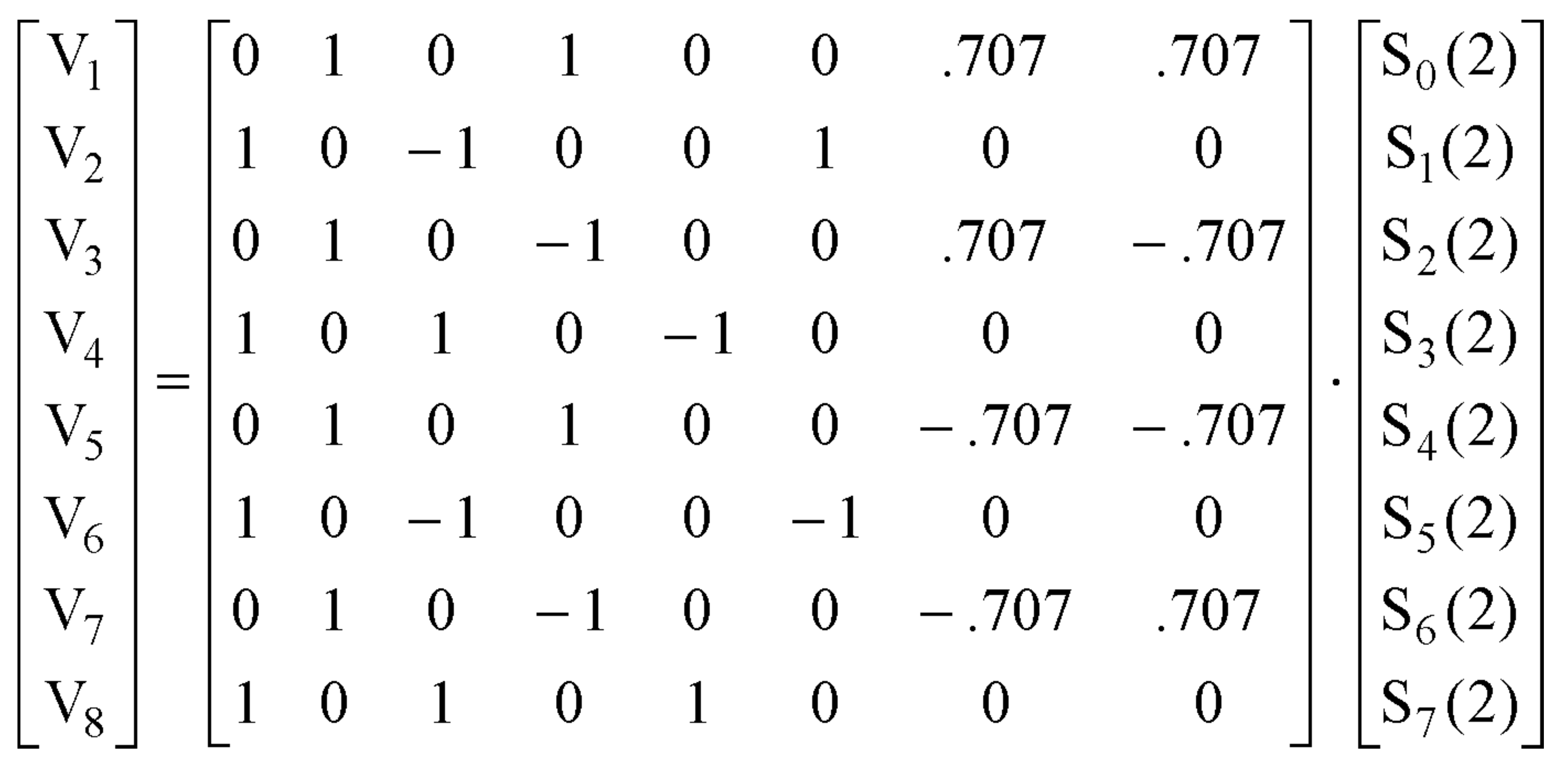}
\end{split}
.
\end{equation*}

\begin{figure}
\centering
\includegraphics[width=6cm]{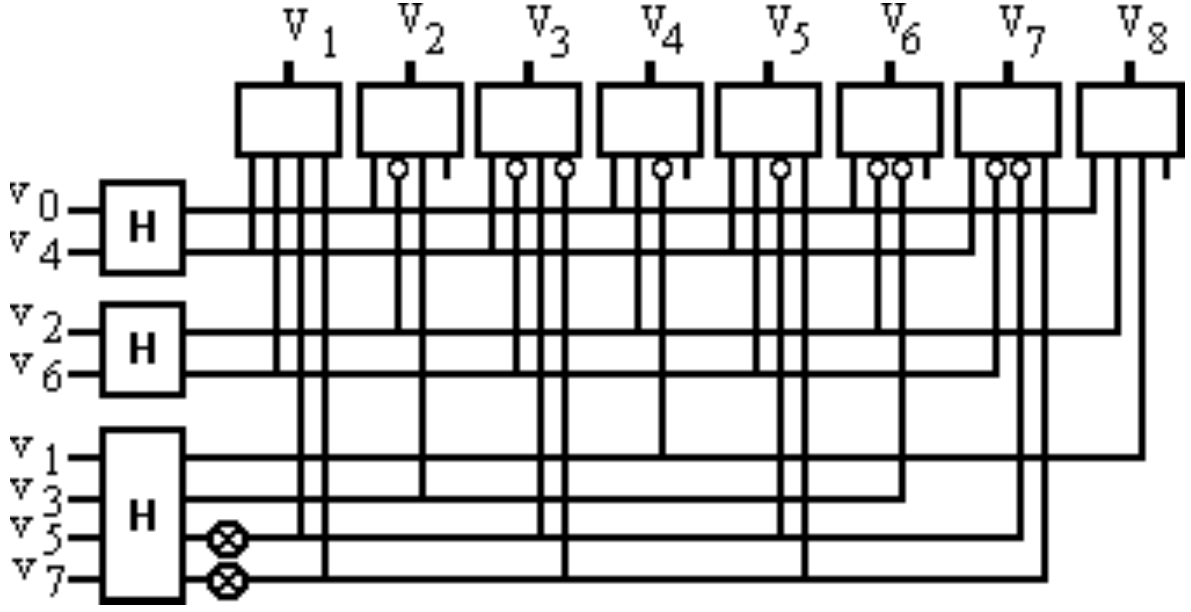}
\caption{The 8-point DHT signal flow diagram.}
\label{figure-3}
\end{figure}

The pre-additions terms can be implemented by Walsh-Hadamard instantiations. 
A scheme for the implementation of the 8-point DHT is shown in Figure~\ref{figure-3},
where
only two multiplications by $\sqrt{2}/2 = 0.707\ldots$ are required.
The algorithm complexity for computing the 8-point
DHT is
22~additions and 2~multiplications.

\section{Computing the 12-point DHT}

The 0-order pre-additions (data) are defined as
$S_i(0)=v_i$, $i=0,1,\ldots,N-1$.
%
%Denoting by
The Hartley spectrum can be computed according to
$\mathbf{V}=\mathbf{T}(0) \cdot \mathbf{S}(0)$,
where 
$\mathbf{T}(0) = \mathbf{H}_{12}$
and
$\mathbf{S}(0)=\begin{bmatrix}S_0(0) & S_1(0) & \cdots & S_{11}(0)\end{bmatrix}^\top$.
Applying the same reasoning of the previous section, 
we define:
\begin{align*}
S_0(1)&=v_0+v_6, \qquad S_1(1)=v_0-v_6, \\
S_2(1)&=v_3+v_9, \qquad S_3(1)=v_3-v_9, \\
S_4(1)&=v_1+v_7, \qquad S_5(1)=v_1-v_7, \\
S_6(1)&=v_2+v_8, \qquad S_7(1)=v_2-v_8, \\
S_8(1)&=v_4+v_{10}, \qquad S_9(1)=v_4-v_{10}, \\
S_{10}(1)&=v_5+v_{11}, \qquad S_{11}(1)=v_5-v_{11}
.
\end{align*}
The resulting transform is:
\begin{equation*}
\begin{split}
\includegraphics[width=11cm]{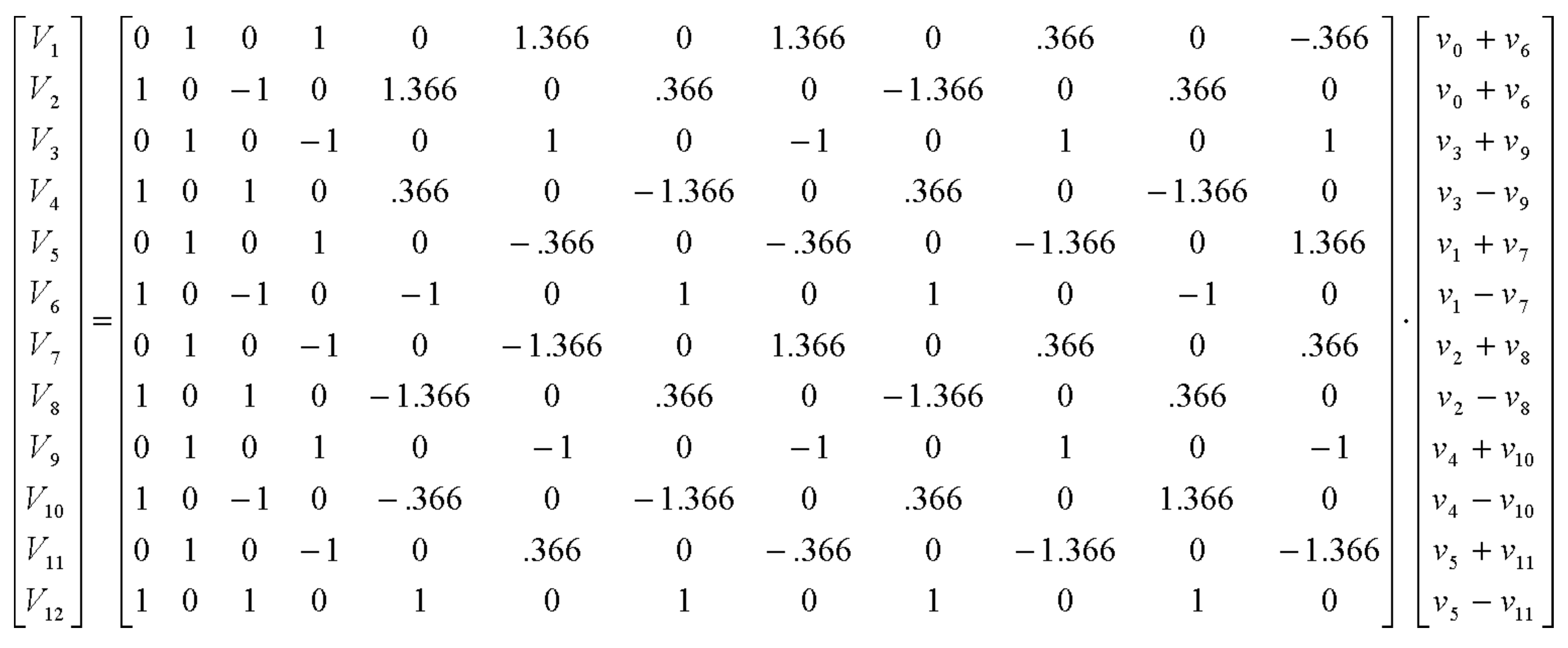}
\end{split}
.
\end{equation*}
Above matrix is denoted as $\mathbf{T}(1)$.
Therefore, this equation can be written as 
$\mathbf{V}= \mathbf{T}(1) \cdot \mathbf{S}(1)$,
where
$\mathbf{S}(1)=\begin{bmatrix}S_0(1) & S_1(1) & \cdots & S_{11}(1)\end{bmatrix}^\top$.
Observing the remaining symmetries, 
we also define
the
2nd-order pre-additions (layer \#2):
\begin{align*}
S_0(2)&=v_0+v_6, S_1(2)=v_0-v_6, 
S_2(2)=v_3+v_9, S_3(2)=v_3-v_9,
\\
S_4(2)&=(v_1+v_7)+(v_4+v_{10}), 
S_5(2)=(v_1+v_7)-(v_4+v_{10}),
\\
S_6(1)&=(v_1-v_7) + (v_2-v_8), 
S_7(2)=(v_1-v_7) - (v_2-v_8),
\\
S_8(2)&=(v_2+v_8) +(v_5+v_{11}), 
S_9(2)=(v_2+v_8) - (v_5+v_{11}),
\\
S_{10}(2)&=(v_4-v10) + (v_5-v_{11}), 
S_{11}(2)=(v_4-v_{10}) - (v_5-v_{11})
.
\end{align*}

We have then:
\begin{equation*}
\begin{split}
\includegraphics[width=12cm]{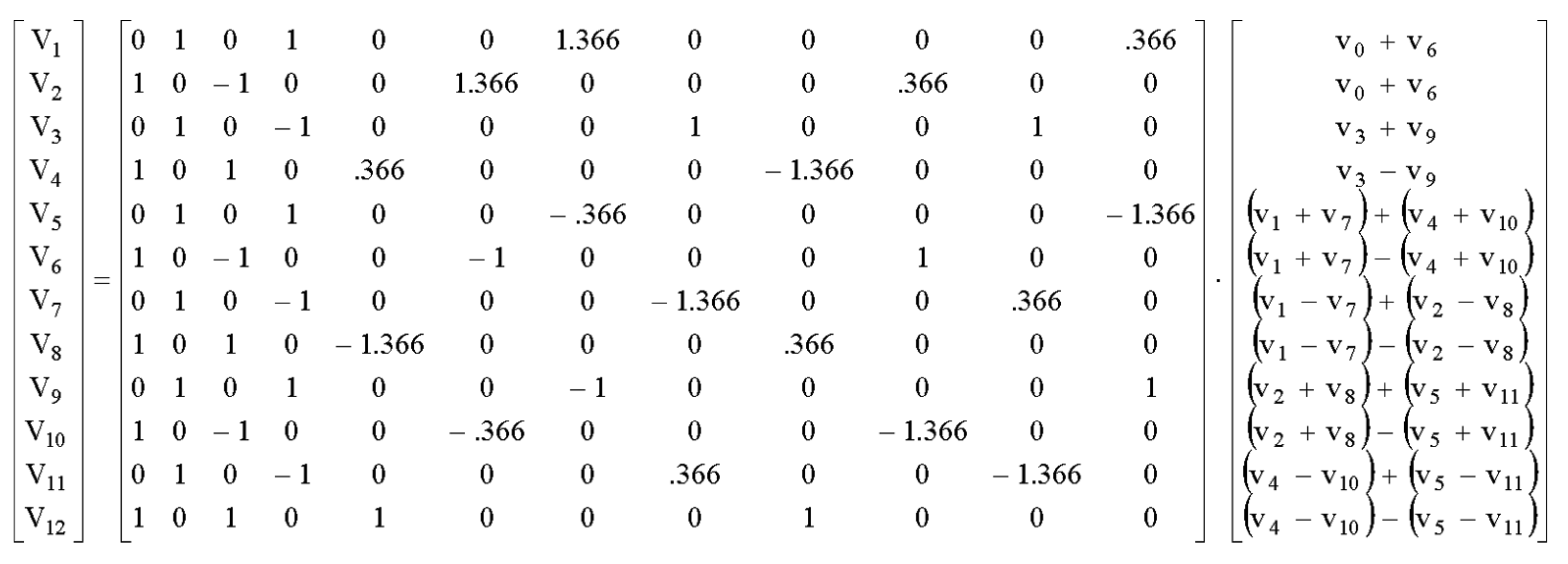}
\end{split}
.
\end{equation*}

The spectrum can be computed in terms of the 2nd layer pre-additions as
$\mathbf{V}=\mathbf{T}(2) \cdot \mathbf{S}(2)$,
where
$\mathbf{T}(2)$ is the 12$\times$12 matrix above
and
$\mathbf{S}(2)=\begin{bmatrix}S_0(2) & S_1(2) & \cdots & S_{11}(2)\end{bmatrix}^\top$.

There is no pair of non-combined identical columns left (signs of elements not considered).
However,
the integer part of the elements greater than unity into the
$\mathbf{T}(2)$ matrix can be handled separately.
Spectral component substitutions to take into account
the special addition to balance the matrix
is shown below:
\[
\begin{array}{lcl}
V_1	& \rightarrow	&[(v_1-v_7)+(v_2-v_8)]=S_6(2)
\\
V_2	& \rightarrow	&[(v_1+v_7)-(v_4+v_10)]=S_5(2)
\\
V_3	& \rightarrow	&0
\\
V_4	& \rightarrow	&-[(v_2+v_8)+(v_5+v_11)]=-S_8(2)
\\
V_5	& \rightarrow	&-[(v_4-v_10)-(v_5-v_11)]=-S_11(2)
\\
V_6	& \rightarrow	&0
\\
V_7	& \rightarrow	&-[(v_1-v_7)-(v_2-v_8)]=-S_7(2)
\\
V_8	& \rightarrow	&-[(v_1+v_7)+(v_4+v_10)]=-S_4(2)
\\
V_9	& \rightarrow	&0
\\
V_{10}	& \rightarrow	&-[(v_2+v_8)-(v_5+v_11)]=-S_9(2)
\\
V_{11}	& \rightarrow	&-[(v_4-v_10)+(v_5-v_11)]=-S_{10}(2)
\\
V_{12}	& \rightarrow	& 0
\end{array}
\]

The procedure of combining pair of columns can be iterated yielding the following new pre-addition sets:
(3rd-order pre-additions (layer \#3))
\begin{align*}
S_0(3)&=v_0+v_6, S_1(3)=v_0-v_6,S_2(3)=v_3+v_9, S_3(3)=v_3-v_9,
\\
S_4(3)&=[(v_1+v_7)+ (v_4+v_{10})] +[(v_2+v_8)  +  (v_5+v_{11})],
\\
S_5(3)&=[(v_1+v_7)+ (v_4+v_{10})] -[(v_2+v_8)  +  (v_5+v_{11})],
\\
S_6(3)&=[(v_1+v_7) - (v_4+v_{10})]+ [(v_2+v_8)   -  (v_5+v_{11})],
\\
S_7(3)&=[(v_1+v_7) - (v_4+v_{10})] - [(v_2+v_8)   -  (v_5+v_{11})],
\\
S_8(3)&=[(v_1-v_7)  - (v_2-v_8)] +[(v_4-v_{10}) +  (v_5-v_{11})],
\\
S_9(3)&=[(v_1-v_7)  - (v_2-v_8)] - [(v_4-v_{10})  +  (v_5-v_{11})],
\\
S_{10}(3)&=[(v_1-v_7) + (v_2-v_8)] + [(v_4-v_{10})  -  (v_5-v_{11})],
\\
S_{11}(3)&=[(v_1-v_7) + (v_2-v_8)] -[(v_4-v_{10})  -  (v_5-v_{11})]
.
\end{align*}
The final relationship between the Hartley spectrum and the pre-additions can be established:
\begin{equation*}
\begin{split}
\includegraphics[width=11cm]{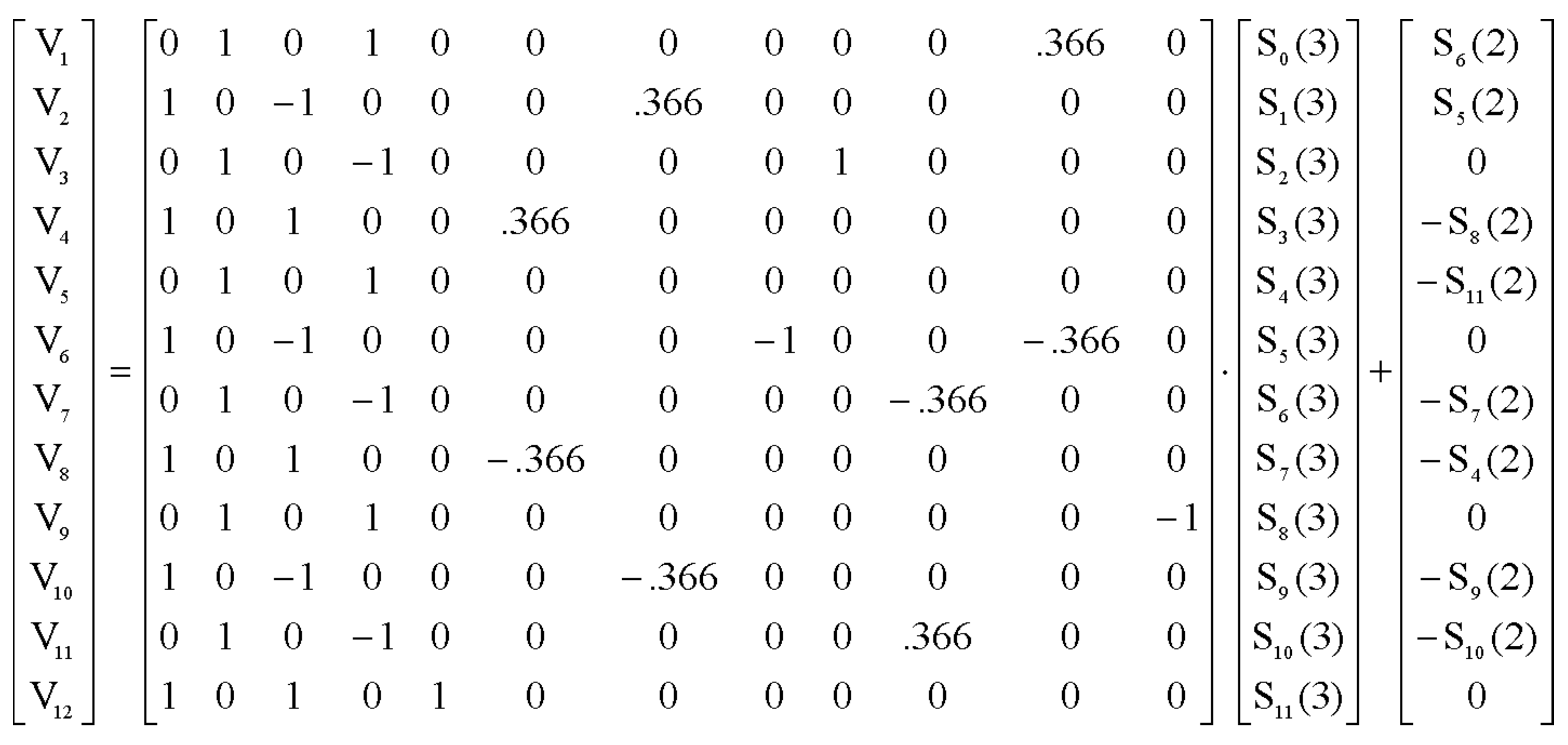}
\end{split}
.
\end{equation*}
The only four real floating-point multiplications required are
$\frac{\sqrt{3}-1}{2} \times [S_5(3), S_6(3), S_9(3), S_{10}(3)]$.
Notice that 
$\frac{\sqrt{3}-1}{2} \approx 0.366\ldots$
The corresponding block diagram is sketched in Figure~\ref{figure-4} below.
The complexity of the suggested implementation is given by 52~additions and 4~multiplications.

\begin{figure}
\centering
\includegraphics[width=6cm]{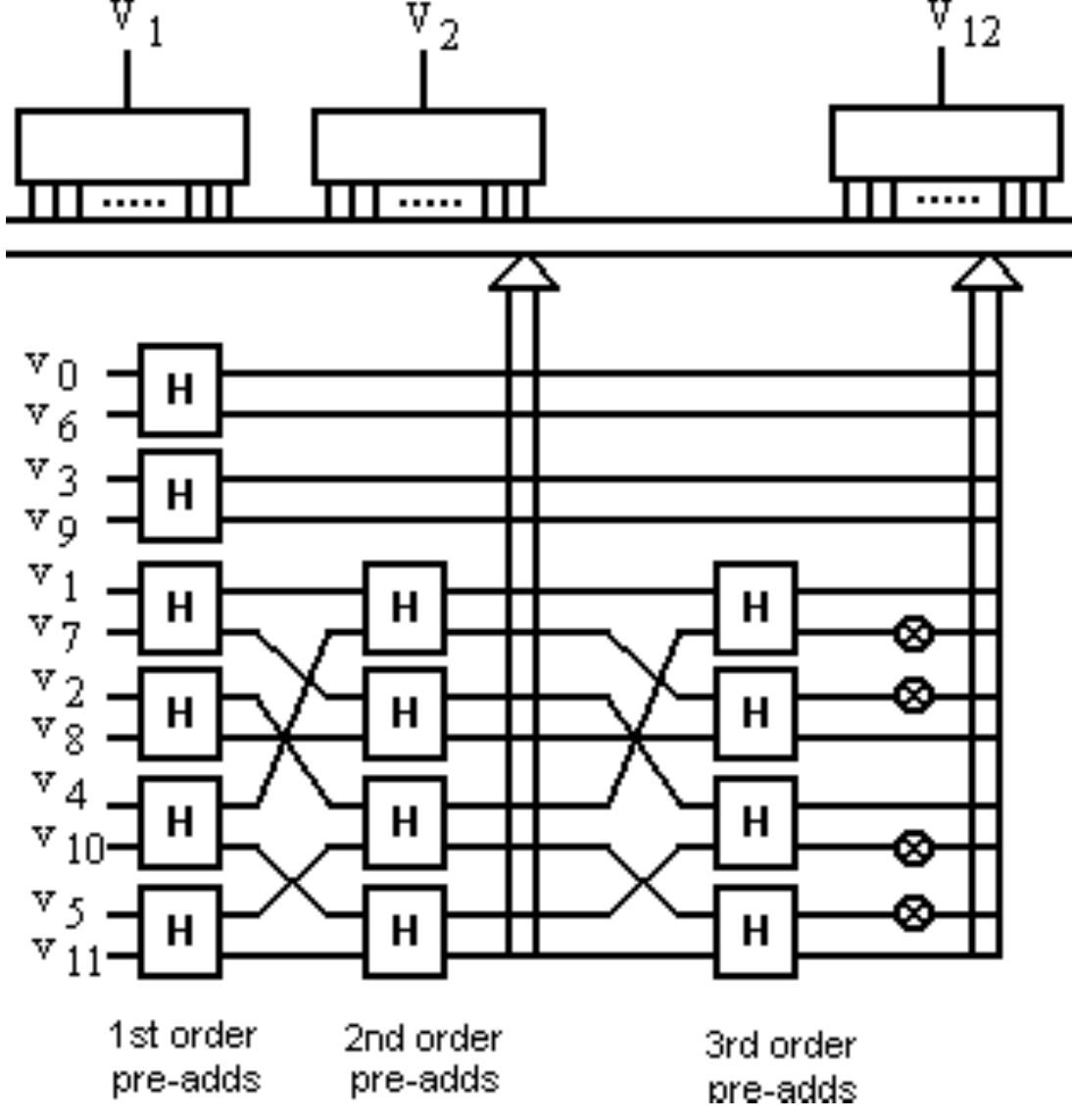}
\caption{The 12-point DHT fast algorithm diagram.}
\label{figure-4}
\end{figure}

\section{Computing the 24-point DHT}

Following the similar steps as before, 
the 0-order pre-additions are defined as 
$S_i(0)=v_i$, $i = 0, 1, \ldots,23$. 
We have the expression below:
\begin{equation*}
\begin{split}
\includegraphics[width=16cm]{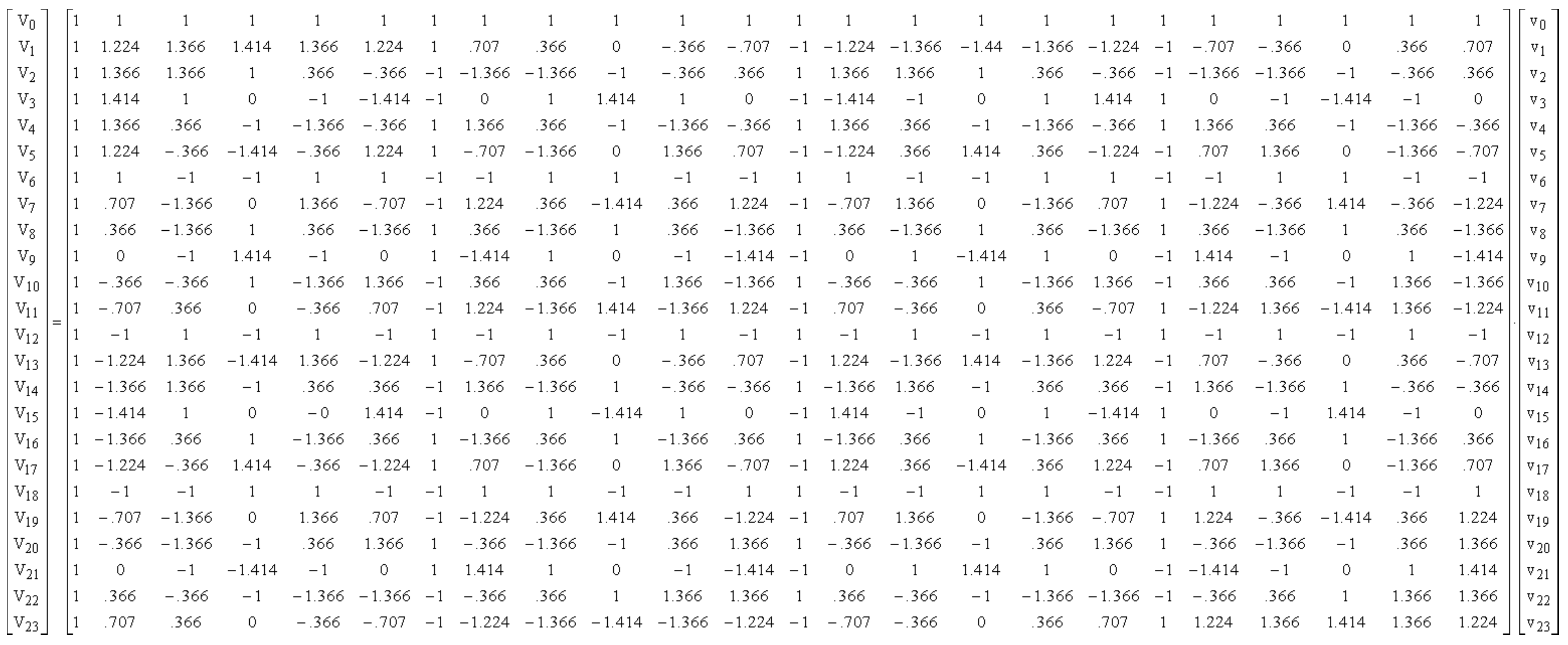}
\end{split}
.
\end{equation*}

Going further, the 1st-order pre-additions (layer \#1) are:
\begin{align*}
S_0(1)&=v_0+v_{12}, S_1(1)=v_0-v_{12}, S_2(1)=v_1+v_{13}, S_3(1)=v_1-v_{13},
\\
S_4(1)&=v_2+v_{14}, S_5(1)=v_2-v_{14}, S_6(1)=v_3+v_{15}, S_7(1)=v_3-v_{15},
\\
S_8(1)&=v_4+v_{16}, S_9(1)=v_4-v_{16}, S_{10}(1)=v_5+v_{17}, S_{11}(1)=v_5-v_{17},
\\
S_{12}(1)&=v_6+v_{18}, S_{13}(1)=v_6-v_{18}, S_{14}(1)=v_7+v_{19}, S_{15}(1)=v_7-v_{19},
\\
S_{16}(1)&=v_8+v_{20}, S_{17}(1)=v_8-v_{20}, S_{18}(1)=v_9+v_{21}, S_{19}(1)=v_9-v_{21},
\\
S_{20}(1)&=v_{10}+v_{22}, S_{21}(1)=v_{10}-v_{22}, S_{22}(1)=v_{11}+v_{23}, S_{23}(1)=v_{11}-v_{23}
.
\end{align*}
A new set of pre-addition can be considered. 
Let the 2nd-order pre-additions be:
\begin{align*}
S_0(2)&=S_0(1), S_1(2)=S_1(1), S_2(2)=S_{12}(1), S_3(2)=S_{13}(1),
\\
S_4(2)&=S_2(1)+S_{14}(1), S_5(2)=S_2(1)-S_{14}(1),
\\
S_6(2)&=S_3(1)+S_{11}(1), S_7(2)=S_3(1)-S_{11}(1), 
\\
S_8(2)&=S_4(1)+S_{16}(1), S_9(2)=S_4(1)-S_{16}(1),
\\
S_{10}(2)&=S_5(1)+S_9(1), S_{11}(2)=S_5(1)-S_9(1),
\\
S_{12}(2)&=S_8(1)+S_{20}(1), S_{13}(2)=S_8(1)-S_{20}(1),
\\
S_{14}(2)&=S_{10}(1)+S_{22}(1), S_{15}(2)=S_{10}(1)-S_{22}(1),
\\
S_{16}(2)&=S_{15}(1)+S_{23}(1), S_{17}(2)=S_{15}(1)-S_{23}(1),
\\
S_{18}(2)&=S_{17}(1)+S_{21}(1), S_{19}(2)=S_{17}(1)-S_{21}(1),
\\
S_{20}(2)&=S_6(1)+S_{18}(1), S_{21}(2)=S_6(1)-S_{18}(1),
\\
S_{22}(2)&=S_7(1)+S_{19}(1), S_{23}(2)=S_7(1)-S_{19}(1)
.
\end{align*}
Again, we have a few cases where the pair do not match perfectly. Applying the same strategy adopted in the 12-blocklength case, we put apart some matrix components in order to ``balance'' the matrix. The 3rd-order pre-additions follows:
\begin{align*}
S_0(3)&=S_0(2), S_1(3)=S_1(2), S_2(3)=S_2(2), S_3(3)=S_3(2), 
\\
S_4(3)&=S_{20}(2), S_5(3)=S_{21}(2), 
\\
S_6(3)&=S_4(2)+S_{12}(2), S_7(3)=S_4(2)-S_{12}(2), 
\\
S_8(3)&=S_5(2)+S_9(2), S_9(3)=S_5(2)-S_9(2),
\\
S_{10}(3)&=S_8(2)+S_{14}(2), S_{11}(3)=S_8(2)-S_{14}(2),
\\
S_{12}(3)&=S_{13}(2)+S_{15}(2), S_{13}(3)=S_{13}(2)-S_{15}(2),
\\
S_{14}(3)&=S_{22}(2)+S_{23}(2), S_{15}(3)=S_{22}(2)-S_{23}(2),
\\
S_{16}(3)&=S_{10}(2)+S_{19}(2), S_{17}(3)=S_{10}(2)-S_{19}(2),
\\
S_{18}(3)&=S_{11}(2)+S_{18}(2), S_{19}(3)=S_{11}(2)-S_{18}(2),
\\
S_{20}(3)&=S_6(2), S_{21}(3)=S_7(2), S_{22}(3)=S_{16}(2), S_{23}(3)=S_{17}(2)
.
\end{align*}

The special addition vector required in this step is written as follows:
\begin{equation*}
\begin{split}
\includegraphics[width=16cm]{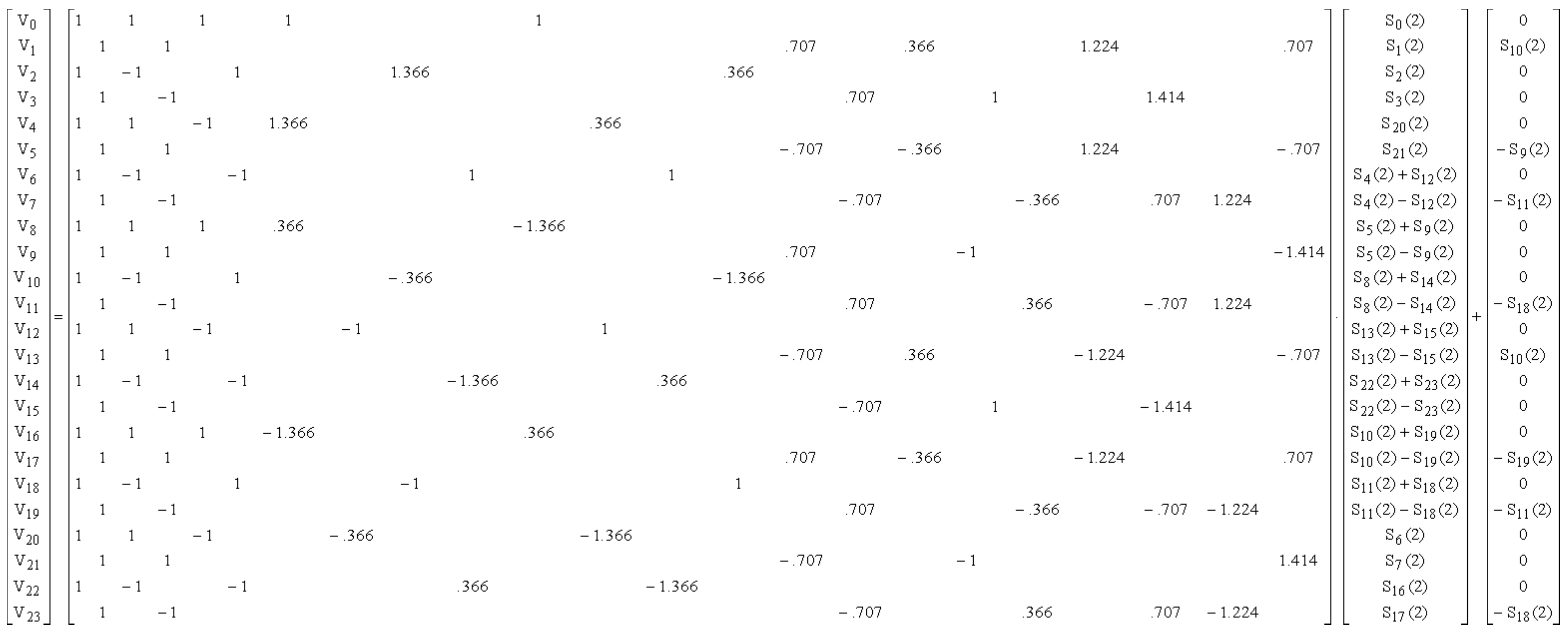}
\end{split}
.
\end{equation*}

The procedure of combining matched columns must be called once more. Making the following definitions, we get the final 4th-order pre-addition, remarking that---as in the previous iteration---another special addition vector must be separated, yielding:
\begin{align*}
S_0(4)&=S_0(3), S_1(4)=S_1(3), S_2(4)=S_2(3), S_3(4)=S_3(3),
\\
S_4(4)&=S_4(3), S_5(4)=S_5(3), S_6(4)=S_{17}(3), S_7(4)=S_{18}(3),
\\
S_8(4)&=S_6(3)+S_{10}(3), S_9(4)=S_8(3)+S_{13}(3), 
\\
S_{10}(4)&=S_8(3)-S_{13}(3), S_{11}(4)=S_6(3)-S_{10}(3),
\\
S_{12}(4)&=S_9(3)+S_{12}(3), S_{13}(4)=S_7(3)+S_{11}(3),
\\
S_{14}(4)&=S_7(3)-S_{11}(3), S_{15}(4)=S_9(3)-S_{12}(3),
\\
S_{16}(4)&=S_{14}(3)+S_{23}(3), S_{17}(4)=S_{14}(3)-S_{23}(3),
\\
S_{18}(4)&=S_{15}(3)+S_{21}(3), S_{19}(4)=S_{15}(3)-S_{21}(3),
\\
S_{20}(4)&=S_{16}(3), S_{21}(4)=S_{19}(3), S_{22}(4)=S_{20}(3), S_{23}(4)=S_{22}(3)
.
\end{align*}

\begin{figure}
\centering
\includegraphics[width=10cm]{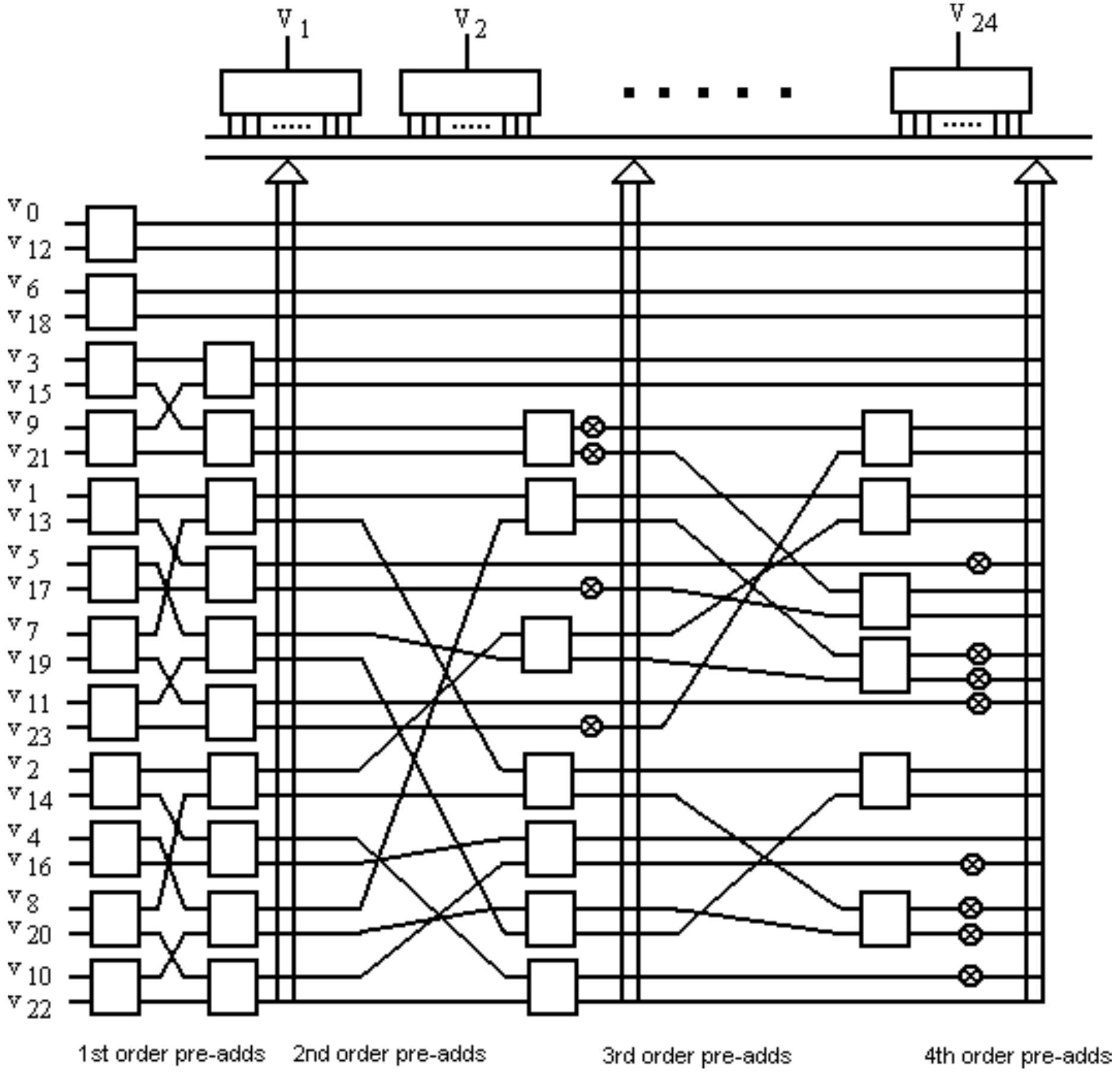}
\caption{The 24-point DHT fast algorithm diagram.}
\label{figure-5}
\end{figure}

Deriving the DHT in terms of the fourth pre-addition layer, we obtain:
\begin{equation*}
\begin{split}
\includegraphics[width=16cm]{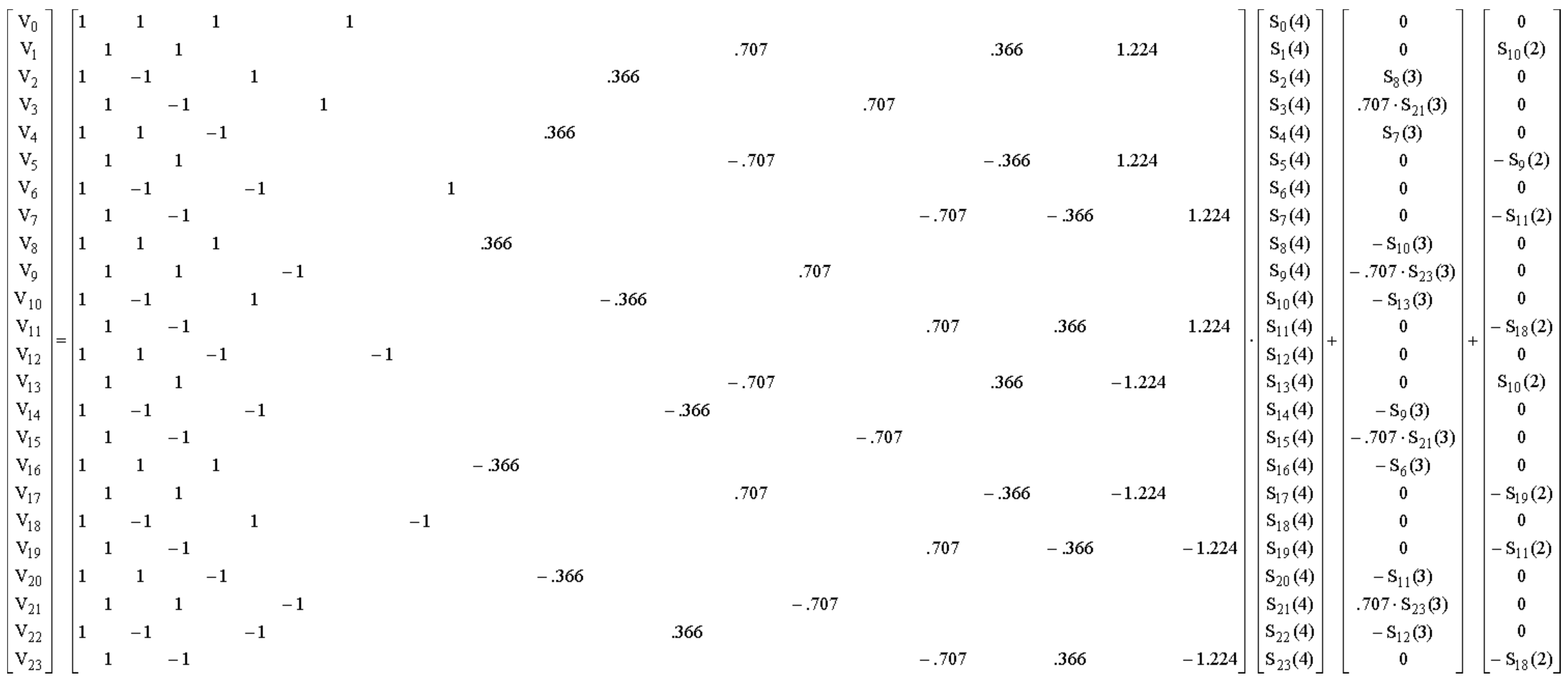}
\end{split}
.
\end{equation*}
Because we have only twelve floating-point multiplication,
the theoretic lower bound on the number of multiplications is achieved.
The corresponding block diagram is depicted in Figure~\ref{figure-5}.
The complexity of the scheme is given by 138~additions and 12~multiplications.

\section{Conclusions}

Fast algorithms for the DHT capable of achieving
the lower bound on the multiplicative complexity of the DFT/DHT
are proposed.
In particular, algorithms 
for short block lengths are presented.
They are based on a multilayer decomposition of the DHT using Walsh-Hadamard transforms.
Each Walsh-Hadamard transfomation implements pre-additions.
These schemes are attractive and easy to implement using in low-cost high-speed dedicated integrated circuits
or
digital signal processors.

{\small
\bibliographystyle{IEEEtran}
\bibliography{ref}
}

\end{document}